\def\oh{\slantfrac{1}{2}}
\def\p{^{\;\prime}}
\def\gev{\text{GeV}}
\def\sp{\text{ }}

\pdfoutput=1
\documentclass[preprint,letter,showpacs,preprintnumbers,amsmath,amssymb,floatfix,nofootinbib]{revtex4}

\usepackage{graphicx}
\usepackage{dcolumn}
\usepackage{bm}

\begin{document}

\title{Effective potentials for heavy quarks above deconfinement}
\author{K. Dusling}
\email{kdusling@quark.phy.bnl.gov}
\author{C. Young}
\email{cyoung@grad.physics.sunysb.edu}
\affiliation{Department of Physics \& Astronomy, State University of New York, Stony Brook, NY 11794-3800, U.S.A.}

\date{\today}  

\begin{abstract}  

Using methods previously developed by Kelbg and others for creating effective 
potentials for electron-ion plasmas, we
investigate quarkonium potentials above deconfinement.  Using 
results for the internal energy of a static quark-antiquark ($Q\bar Q$) pair,
a set of effective potentials are constructed that take into account 
quantum effects and symmetry conditions.  Bound state effects are 
explicitly included in order to account for the strongly coupled nature of 
the plasma.  It is hoped that these effective potentials will be useful in 
simulations of heavy quarks or plasma quasiparticles when the dynamics is treated classically.

\end{abstract}


\maketitle  

\section{Introduction}

\label{intro}

It has recently been noted that the matter produced in heavy ion 
collisions at RHIC cannot be weakly coupled but instead behaves as a good 
liquid \cite{sqgp1, sqgp2}.  The evidence for the sQGP is large but mainly 
consists of the following two points: 1. the observed collective flows at 
RHIC can be explained by hydrodynamics showing that the dissipative lengths 
are very short and 2. binary bound states are seen to exist in lattice 
simulations above $T_c$ and are also predicted \cite{sqgp3} using lattice 
interparticle potentials. 

Since a complete quantum description tends to rely on perturbative methods 
other approaches have to be adopted in order to perform calculations at 
strong coupling. One approach as was discussed in \cite{MD1, MD2,MD3}, is 
to model the strongly interacting quark and gluon quasiparticles as a 
classical non-relativistic colored Coulomb gas. This model is analyzed using 
molecular dynamics (MD) simulations in which real time correlators can be 
extracted. The interaction between quarks was taken as a Coulomb potential 
with a strong repulsive core in order to {\em mock up} quantum-mechanical 
effects.  

A second example is a model treating charmonium in the sQGP \cite{charmonium},
where initial ensembles of quark pairs from pQCD events are evolved 
classically according to a Langevin model with an additional interaction 
potential included between quark-antiquark pairs. The interaction potential 
was taken from lattice $c \bar c$ internal energy data with quantum-mechanical 
effects mimicked by simply turning off the potential below the approximate 
Bohr radius.

If these models are to be refined it is necessary that we consider 
quantum-mechanical effects more carefully. The construction of an effective interaction which mimics the effects of quantum dynamics was 
first performed by Kelbg \cite{Kelbg}. The goal of this work is to 
generate classical effective $q\bar q$ potentials which take into account 
quantum-mechanical effects, first the dynamics and then the symmetry or anti-symmetry which must exist in the wavefunction for the 
$q \bar{q}$ pair. We want to 
find an effective potential $V_{eff}(\vec r,T)$ whose $classical$ Boltzmann 
factor yields the diagonal term of the $quantum-mechanical$ density matrix:
\begin{equation}
\langle \vec r|e^{-\beta \hat{H}}|\vec r\rangle=e^{-\beta V_{eff}(\vec r,T)}\langle \vec r|e^{-\beta\frac{\hat{p}^2}{2m}}|\vec r\rangle
\end{equation}
For example, if this corrected interaction were used in the modeling of 
charmonium discussed in the previous paragraph, the final equilibrium 
distribution obtained by the Fokker-Planck evolution would be the correct 
quantum-mechanical distribution. This would obviously improve estimates 
for the intermediate states as well.  These effective interactions would 
also have use in the MD simulations discussed earlier.

In this work we calculate effective quark-antiquark potentials 
from the internal energy of static $Q \bar Q$ pairs 
calculated in \cite{kaczmarek} based on lattice calculations of the $Q \bar Q$ $free$ 
energy in \cite{Kaczmarek2}. Section \ref{veff} of this paper summarizes a 
number of previous works~\cite{Hilton, bpl, corrcoulomb, WDK}, including the definition of the 
Slater sum and its use in creating effective potentials. The original 
results of this paper start in section \ref{veff_lat}, where the methods 
presented in section \ref{veff} are applied to potentials extracted from 
lattice data.

\section{Creating Effective Potentials}

\label{veff}

\subsection{The Slater Sum}

\label{slater}

Anticipating applications of this work to many-particle systems, we 
define the $N$-particle Slater sum:

\begin{equation}
S^{(N)}(\vec{r}_1,...,\vec{r}_N)=N!\Lambda^{3N}\sum_i |\Psi_i(\vec{r}_1,...,
\vec{r}_N)|^2e^{-\beta E_i}
\end{equation}

where $\Psi_i$ are the N-particle energy eigenfunctions and 
$\Lambda=\sqrt{\frac{h^2}{2\pi m k T}}$ is the thermal wavelength. 
Integrating the Slater sum in coordinate space yields the partition function:

\begin{equation}
Z=\frac{1}{N!\Lambda^{3N}}\int (d{\vec r}_1...d{\vec r}_N)S^{(N)}(\vec{r}_1
,...,\vec{r}_N)
\label{pf}
\end{equation}

In the case of ideal particles using Boltzmann statistics the partition 
function is given as $Z_{MB}=\frac{V^N}{N!\Lambda^{3N}}$.  The corresponding 
free energy can then be separated in the following way:

\begin{equation}
F=-k_B T \ln Z=F_{MB}-k_B T \ln\frac{Z}{Z_{MB}}
\end{equation}

Substituting eq.~\ref{pf} for $Z$ in the above equation yields:

\begin{equation}
F=F_{MB}-k_B T \ln\frac{1}{V^N}\int (d{\vec r}_1...d{\vec r}_N)S^{(N)}
(\vec{r}_1,...,\vec{r}_N)
\end{equation}

This expression should now be compared to the classical counterpart of the 
free energy:

\begin{equation}
F_{cl}=F_{MB}-k_B T \ln\frac{1}{V^N}\int (d{\vec r}_1,...,d{\vec r}_N) 
e^{-\beta\sum_{i<j}V_{ij}}
\end{equation}

In the classical limit one obtains:
\begin{equation}
S_{cl}^{(N)}(\vec{r}_1,...,\vec{r}_N)=e^{-\beta\sum_{i<j}V_{ij}}
\end{equation}
and near the classical limit one can include multi-body interactions due to 
the quantum correlations between particles:
\begin{equation}
S^{(N)}(\vec{r}_1,...,\vec{r}_N)=e^{-\beta\sum_{i<j}u_{ij}-\beta\sum_{i<j<k}
u_{ijk}+...}
\end{equation}
where $u_{ij}$ and $u_{ijk}$ are effective two- and three-body interactions 
respectively and can be expressed in terms of two- and three-particle Slater 
sums as:

\begin{align}
u_{ij}=-k_B T \ln S^{(2)}(r_i,r_j)\nonumber\\
u_{ijk}=-k_B T \ln \frac{S^{(3)}(r_i,r_j,r_k)}{S^{(2)}(r_i,r_j)S^{(2)}
(r_i,r_k)S^{(2)}(r_j,r_k)}
\label{eq:effpot}
\end{align}

Keeping only the first term, $u_{ij}$, in the series for $S^{(N)}$ accounts for quantum effects up to first order in the coupling parameter, $\Gamma=\frac{(Ze)^2}{a_{WS}T}$ where $Ze, a_{WS}, T$ are respectively the ion charge, the Wigner-Seitz radius $a_{WS}=(3/4\pi n)^{1/3}$ and the temperature.  The second term $u_{ijk}$ introduces corrections to the effective potential of order $\Gamma^2$.

 \subsection{Effective potentials in the binary approximation}

\label{2veff}

The previous section motivates us to take the binary approximation:
\begin{equation}
S^{(N)} \approx e^{-\beta \sum_{i<j} u_{ij}}
\end{equation}
where $u_{ij}$ is defined as in eq.~\ref{eq:effpot}. This approximation 
is exact for a 2-body systems (for example the model discussed in section 
\ref{intro} where a heavy quark only interacts with its diagonal partner). 
For the remainder of this paper, we now focus on an explicit calculation of 
$u_{ij}$. Our method involves examining the two-particle density matrix:

\begin{equation}
\rho(\vec{r}_1,\vec{r}_2,\vec{r}_1\p,\vec{r}\p_2)=
\sum_i \Psi^*_i(\vec{r}_1,\vec{r}_2)\Psi_i(\vec{r}\p_1,\vec{r}\p_2)
e^{-\beta E_i}
\end{equation}

where the diagonal term is proportional to the needed Slater sum.  The 
two-particle density matrix can then be factored into center-of-mass and 
relative components, $\rho(\vec{r}_1,\vec{r}_2,\vec{r}\p_1,\vec{r}\p_2)=
\rho(\vec{R},\vec{R}\p)\rho(\vec{r},\vec{r}\p)$.  

The two-body effective potential can then be found in the limit as 
$\vec{R}\p\to\vec{R}$ and $\vec{r}\p\to\vec{r}$. Since 
$\rho(\vec{R},\vec{R})=1$ one is left with solving for 
$\rho(\vec{r},\vec{r})$.

Acting on $\rho(\vec{r},\vec{r}\p)$ with the one-particle Hamiltonian\footnote{From here on we use standard high-energy units where $\hbar=c=1$}, 
$H=-\frac{1}{2m}\nabla^2+V(\vec{r})$ where $m^{-1}=m_1^{-1}+m_2^{-1}$ is the reduced 
mass of the two particle system and $\vec{r}$ is the relative coordinate it 
can be shown that $\rho$ satisfies:
\begin{equation}
H(\vec{r})\rho(\vec{r}, \vec{r}^{'}, \beta)=-\frac{\partial \rho}{\partial \beta}
\label{eq:bloch}
\end{equation}

The boundary condition on the density matrix is given from the fact that 
\{$\Psi_i$\} is a complete basis:
$\rho(\vec{r},\vec{r}\p;\beta=0)=\delta(\vec{r}-\vec{r}\p)$

We use the same method of solution on eq. \ref{eq:bloch} which is given 
in \cite{Hilton}.  The solution to the free-particle density matrix (V=0 in eq.~\ref{eq:bloch}) is 
given as
\begin{equation}
\rho_0(\vec{r},\vec{r}\p,\beta)=\frac{1}{(\sqrt{4\pi}\lambda)^3}
\exp( \frac{-|\vec{r}-\vec{r}\p|^2}{4\lambda^2})
\label{eq:fp}
\end{equation} 
where $\lambda=\sqrt{\frac{1}{2mT}}$ and $m$ is the reduced mass of the 
two-particle system.

We then define the effective potential as in the introduction:
\begin{equation}
\rho(\vec{r},\vec{r}\p,\beta)=\rho_0 e^{-\beta u(\vec{r},\vec{r}\p,\beta)}
\label{eq:general}
\end{equation}
The effective two body potential $u_{ij}$ of interest is simply the diagonal 
part of $u(\vec r,\vec r\p,\beta)$ when $\vec r\p=\vec r$. 

Substituting the above form for the density matrix \ref{eq:general} into the Bloch equation 
\ref{eq:bloch} 
one finds that $u(\vec r,\vec r\p,\beta)$ satisfies:
\begin{equation}
\frac{1}{2m}\beta\nabla^2 u-\frac{1}{2m}\beta^2(\nabla u)^2
-(\vec r -\vec r\p)\cdot\nabla u+V(\vec r)=
\beta(\partial u/\partial \beta)+u
\label{eq:exact}
\end{equation}

In the limit that $u$, $\nabla_r u$, or $\beta$ is small, eq. \ref{eq:exact} 
may be linearized:
\begin{equation}
\frac{1}{2m}\beta\nabla^2 u_1
-(\vec r -\vec r\p)\cdot\nabla u_1+V(\vec r)=
\beta(\partial u_1/\partial \beta)+u_1
\label{eq:approx}
\end{equation}
the solution of which we call $u_1(\vec r,\vec r\p,\beta)$. As has been 
emphasized in \cite{Hilton}, this condition is far more permissive 
than the typical condition from perturbation theory that $u$ be small.
This equation can be solved exactly:
\begin{equation}
u_1(\vec r,\vec r\p)=\int d\vec r_1 G(\vec r,\vec r\p,\vec r_1,\beta)V(\vec r_1)
\label{eq:u1e}
\end{equation}
where the Green function $G$ is given by:
\begin{equation}
G(\vec r,\vec r\p,\vec r_1,\beta)=\frac{m}{2\pi\beta} \exp(-\frac{m|\vec r-\vec r\p|^2}{2\beta})\Bigl[ \frac{|\vec r-\vec r_1|-|\vec r\p-\vec r_1|}{|\vec r -\vec r_1||\vec r\p-\vec r_1|}\Bigr] \exp(-\frac{m(|\vec r-\vec r_1|+|\vec r\p-\vec r_1|)^2}{2\beta})
\label{eq:green}
\end{equation}
For a spherically symmetric potential $V(\vec r)$, the angular integration in 
the above equation can be performed and the final result for the diagonal 
term is given as $u_1(\vec{r}=\vec{r}\p,\beta)$ is:
\begin{equation}
u_1(\vec{r},\beta)=\frac{1}{\lambda}\sqrt{\frac{\pi}{4}}\int_0^\infty\frac{r_1}
{r}V(r_1)
\biglb[
\text{erf}(\frac{|r+r_1|}{\lambda}) - \text{erf}(\frac{|r-r_1|}{\lambda} 
\biglb]dr_1
\label{eq:u1}
\end{equation}

Equation \ref{eq:u1} will be the starting point for our future analysis of 
lattice potentials.  It can be shown that the same result can be found by 
solving for the Slater sum $S^{(2)}$ explicitly using plane waves for the 
wavefunction and solving perturbatively in the coupling constant.  

\subsubsection{Bound States}

It has already been shown in \cite{Hilton} that the linearized approximation which yielded eq. 
\ref{eq:approx} is satisfactory assuming the contribution from bound states 
is small.  The term dropped from the linearized equation, $\frac{1}{2}\beta^2(\nabla u)^2$,  becomes  
significant when the gradient of the effective potential is on the order of the temperature.  This is exactly the condition leading to large contributions from  
the lowest bound states to the density matrix. 
In order to take into account the effect of bound 
states explicitly we follow \cite{bpl}, where the bound and free states are considered 
as two separate contributions to the Slater sum.  Both the partition function and the 
two-particle Slater sum can be separated as
\begin{equation}
S^{(2)}(r,\beta)=(1-P\p)S^{(2)}+P\p S^{(2)},
\end{equation}
where the operator $P\p$ projects out or removes the free component.  The terms on the right-hand side correspond to the free and bound state contribution to the Slater sum, respectively.

There are various ways of defining the projection operator $P\p$, so that it separates the Slater sum 
into a part dominated by the free contribution (call it $S^{(2)}_f$) and a part dominated by the 
bound state contribution. In principle, these are merely different conventions, however since we 
will identify $S^{(2)}_f$ with the Slater sum we already calculated, some conventions are better than 
others.
Explicitly, one can use the Riewe-Rompe convention which acts as a sharp cutoff between the bound and continuum states.  More specifically, the projection operators acting on the two particle Slater sum yield 
\begin{equation}
S^{(2)}(r,\beta)=S^{(2)}_f(r,\beta)+\sum_{E_i < \epsilon^*} |\Psi_i(r)|^2 e^{-\beta E_i},
\end{equation}
where $\epsilon^*\sim T$.  
Using this convention, we would be adding a term to the Slater sum even where 
$\beta$ is small, where our calculation needs no correction. In order to circumvent this, we use the Brillouin-Planck-Larkin (BPL) convention since it has a continuous transition at the continuum edge,
\begin{equation}
S^{(2)}(r,\beta)=S^{(2)}_f(r,\beta)+\sum_i |\Psi_i(r)|^2[ e^{-\beta E_i} -1+\beta E_i].
\label{eq:bound}
\end{equation}
The advantage of this convention can be seen by examining small and large $\beta$.
When $\beta$ is small, the coefficient $e^{-\beta E_i}-1+\beta E_i \approx (1-\beta E_i) -1+\beta E_i
\approx 0$, and no significant contribution is added to the free Slater sum.  When 
$\beta$ is large, $e^{-\beta E_i}-1+\beta E_i \approx e^{-\beta E_i}$ ($E_i$ is negative), making 
the bound state part the dominant contribution to the Slater sum, as it should be at low temperatures.

In the above expressions $S^{(2)}_f(r,\beta)$ is the free-particle Slater sum and can be found 
by substituting the result from eq. \ref{eq:u1} into eq. \ref{eq:effpot}. 
The second term includes a sum over bound states with wavefunctions given as solutions to the 
Schr\"odinger equation.

\subsubsection{Symmetry Considerations}

We make one final consideration in creating effective potentials: symmetry 
conditions for identical particles. In the case of the identical fermions, 
the two-particle density matrix, written as an imaginary--time propagator, 
becomes
\begin{equation}
\rho(\vec r_1,\vec r_2,\vec r_1\p,\vec r_2\p,\beta )
=\frac{1}{2}(\langle \vec r_1,\vec r_2|-\langle \vec r_2,\vec r_1|)
\exp(-\beta\hat{H})(|\vec r_1\p,\vec r_2\p\rangle-|\vec r_2\p,\vec r_1\p
\rangle)
\end{equation}
For our situation where two particles interact according to a potential 
only dependent on $|\vec r|$, their relative separation, this formula 
dramatically simplifies:
\begin{equation}
\rho(\vec r_1,\vec r_2,\vec r_1\p,\vec r_2\p,\beta)
=\rho_{abs}(\vec R,\vec R\p,\beta)\bigl[\rho_{rel}(\vec r,\vec r\p,\beta)
-\rho_{rel}(\vec r,-\vec r\p,\beta)\bigr]
\label{eq:antisym}
\end{equation}
Again, we will be interested in the limit $\vec r\p=\vec r$.
We may calculate explicitly the off-diagonal term of the density matrix 
needed in 
eq.~\ref{eq:antisym}, or we may follow the approach in \cite{corrcoulomb}
and approximate the off-diagonal term using only the off diagonal free 
particle density matrix
\begin{equation}
\rho_{rel}(\vec r,-\vec r,\beta)\approx \exp(-2m|\vec r|^2/\beta)\rho_{rel}(\vec r,\vec r,\beta)
\end{equation}

The final form for the correctly symmetrized density matrix $\rho_{sym}$ is given as:
\begin{equation}
\label{eq:symf}
\rho_{sym}(\vec{r},\beta)=\frac{1}{\lambda^3}\bigl[1\pm\exp(-2m|\vec r|^2/\beta)\bigr]e^{-\beta u_1(\vec{r})}
\end{equation}
where a minus sign is used for fermions and a plus sign is for bosons.

\subsection{Coulomb Potential}

\label{sec:coulomb}

We now apply the methods outlined in the previous section to the case of 
the Coulomb potential.  We first note that equation~\ref{eq:u1} can be integrated exactly in the case of a Coulomb potential, $V(r)=-1/r$ with the result:

\begin{equation}
u_1(r,\beta)=\frac{-1}{r}\biglb( 1-e^{-r^2/\lambda^2} +\frac{\sqrt{\pi}r}{\lambda}\biglb[1-\text{erf}(\frac{r}{\lambda})\bigrb]\biglb)
\end{equation}

\begin{figure}
\includegraphics[scale=.75]{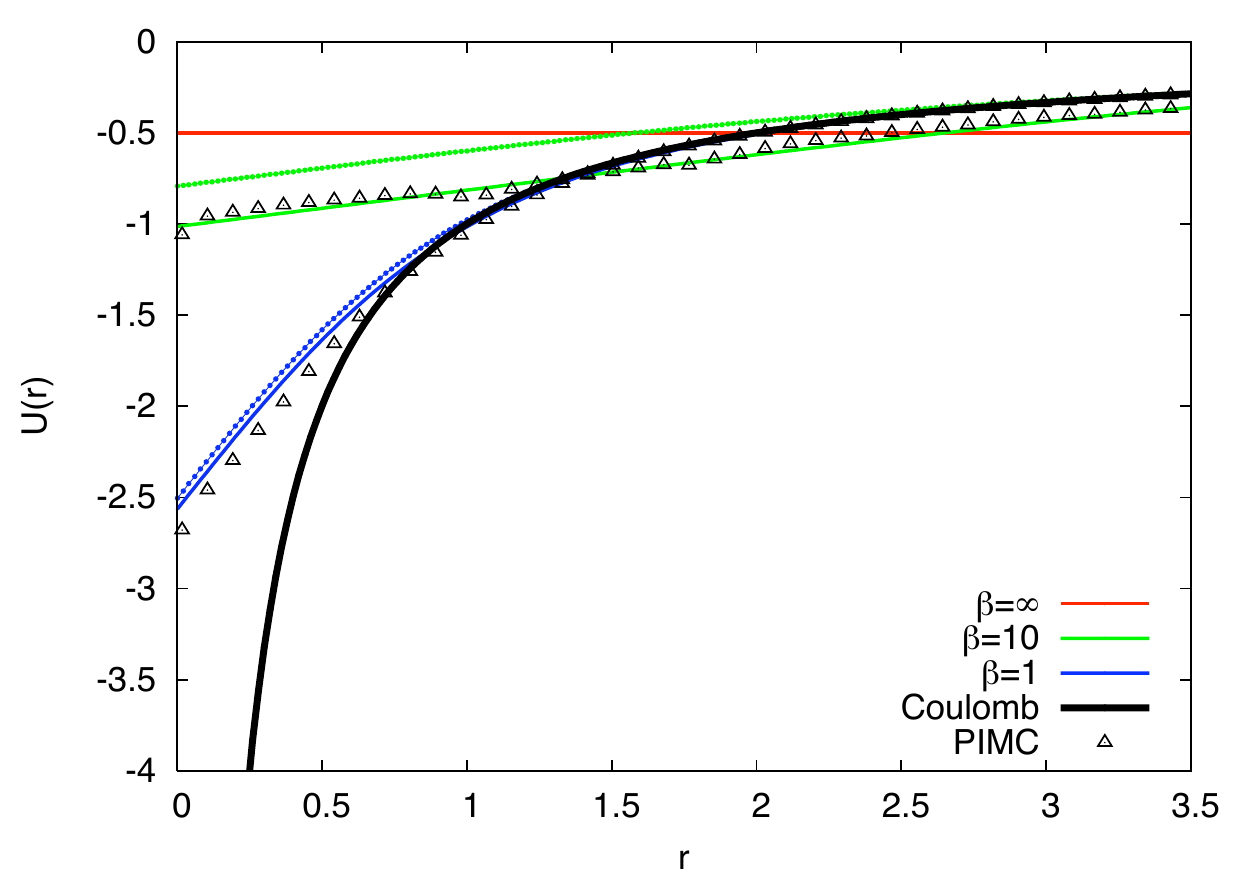}
\caption{Effective Coulomb potential for $\beta=1, 10$.  Dotted lines show 
the Kelbg potential $u_1(r)$ and corresponding solid line shows potential 
including correction from lowest bound state.  The triangles are the results 
of PIMC (see text).}
\label{fig:cbound}
\end{figure}

In figure \ref{fig:cbound} we show the results of including bound states for 
the case of the Coulomb potential.  The dotted green and blue curves show the 
results for $u_1(r)$ for values of $\beta=10$ and $1$ 
respectively. The corresponding solid curves show the results of including 
the first 
lowest bound state in the sum in eqn~\ref{eq:bound}.  The solid red curve 
shows the result at $\beta=\infty$ where the result $u_1$ vanishes but the 
effective potential sits at the energy of the lowest bound state $E=-\oh$ 
in atomic units.

Also shown on this plot (black triangles) are numerical results of a 
Monte-Carlo simulation 
of the density matrix. Paths in imaginary time were sampled from the 
free-particle distribution by way of the Levy construction, and the 
actions of these paths were calculated and averaged according to their 
weights determined from the sampled, free-particle distribution of paths. 
See Ceperley \cite{ceperley} for an extremely useful and pedagogical 
introduction into Monte-Carlo methods for calculating $N$-particle (Bose) 
density matrices.

\begin{figure}
\includegraphics[scale=.75]{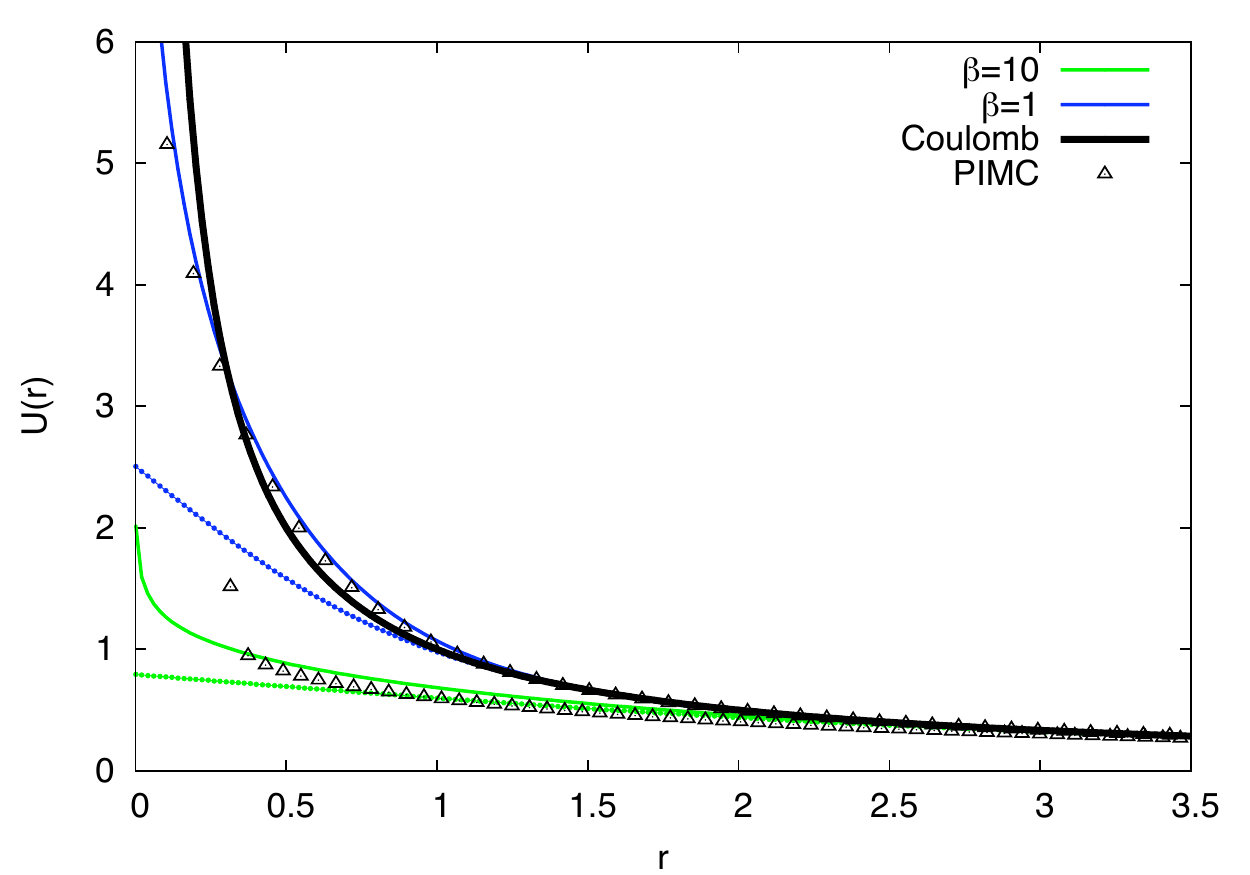}
\caption{Dotted lines show the Kelbg potential for a repulsive electron pair. 
The solid lines show the potential after anti-symmetrization of the density 
matrix.  The triangles are the results of the PIMC.}
\label{fig:symf}
\end{figure}

In fig~\ref{fig:symf} we show the result for two identical electrons after 
the correct symmetrization is preformed.  Again the dotted lines show the 
result of $u_1$ and the corresponding solids lines are the result after 
anti-symmetrization of the density matrix \ref{eq:symf}.

\section{Effective Lattice Potentials}

\label{veff_lat}

\subsection{Lattice Parameterization}

Now we focus on the problem at hand, obtaining an effective potential appropriate for simulations of heavy 
flavor quarks or plasma quasiparticles. Internal and free energies for static quark-antiquark pairs have been 
calculated at temperatures above deconfinement \cite{kaczmarek, Kaczmarek2}, and either of these energies could be used 
to construct an effective interaction. However, it is not clear which potential would best be used in the simulation 
of dynamical heavy quarks. Shuryak and Zahed \cite{Shuryak:2004tx} have argued that the rapid rotation of the quarks in bound states 
causes the timescales of interest to be relatively short, and therefore the appropriate energy for the interaction 
would be the internal energy. However, Mocsy and Petreczky \cite{Mocsy:2005qw}, have found that 
potential models which use the free energy instead of the internal energy as the interaction 
potential best describe the spectral functions which were extracted from the Euclidean correlators on the lattice 
using the maximal entropy method. While this does not yet rule out the argument of Shuryak and 
Zahed (there is still significant uncertainty in the spectral function obtained using the 
maximal entropy method), it clearly establishes that at the time of this publication, there is no clear 
answer concerning which potential best describes the interaction between heavy quarks.

With no specific bias towards the use of either energy for describing the interaction, we will consider the internal energy.
We parameterize the internal energy in the temperature range $1.1-2 T_C$ as follows:

\begin{equation}
\label{eq:param}
E_1(r,T)=-\frac{\alpha(T)}{r}e^{-\mu(T)r}+\frac{\sigma(T)}{b}\tanh(a\mu(T)r)
\end{equation}

This potential is similar in form to the screened Cornell potential as first 
proposed in \cite{Cornell}.  The first part of the potential is the usual 
screened Coulomb potential with temperature dependent coupling ($\alpha$) and 
Debye mass ($\mu$).  These values were extracted from the lattice data and 
parameterized as follows.  For the coupling constant we use a form similar to 
that from the leading log order renormalization group equation, $\alpha(T)=[2.4\ln(2T/T_c)]^{-1}$.  The 
fit to the Debye mass goes as $\mu\propto T$ which is consistent with the 
lowest-order perturbative calculations \cite{Bellac} with additional 
corrections to account for the data better: 
$\mu(T)=0.0675 (T/T_c)-(\frac{0.196T}{T_c})^5$.

The second part of the potential takes on the same form as a screened linear 
confining potential.  We choose a different functional form than the 
usual $\sigma r e^{-\mu r}$ in order that the potential remains flat at 
intermediate and large distances instead of having a maximum and then 
becoming repulsive at larger $r$ as in the case of the Cornell potential. 
We find that a parameterization of the {\em string tension} as 
$\sigma(T)/(\gev^2)=0.32(T_c/T)+0.8(T_c/T)^{16}+0.005$ reproduces the 
lattice data reasonably well across a large range of temperatures.  The 
final parameters are $a=5.5, b=0.37\sp\gev$ and $T_c=0.27\sp\gev$.

The potential we use is then given as:
\begin{equation}
\label{eq:pot}
V_1(T,r)=E_1(T,r)-E_1(T,\infty)
\end{equation}

\begin{figure}
\centerline{\hbox{
\includegraphics[scale=.75]{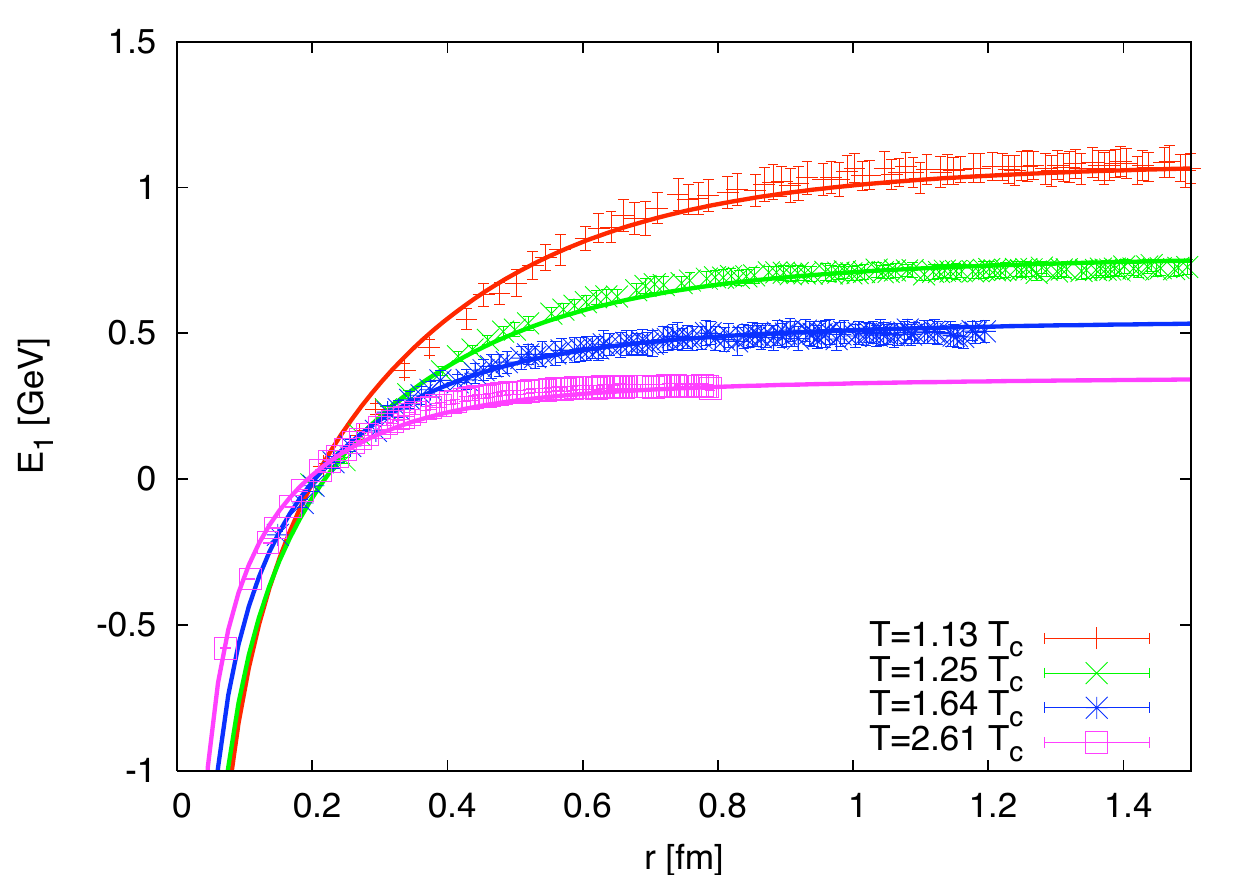}
\includegraphics[scale=.75]{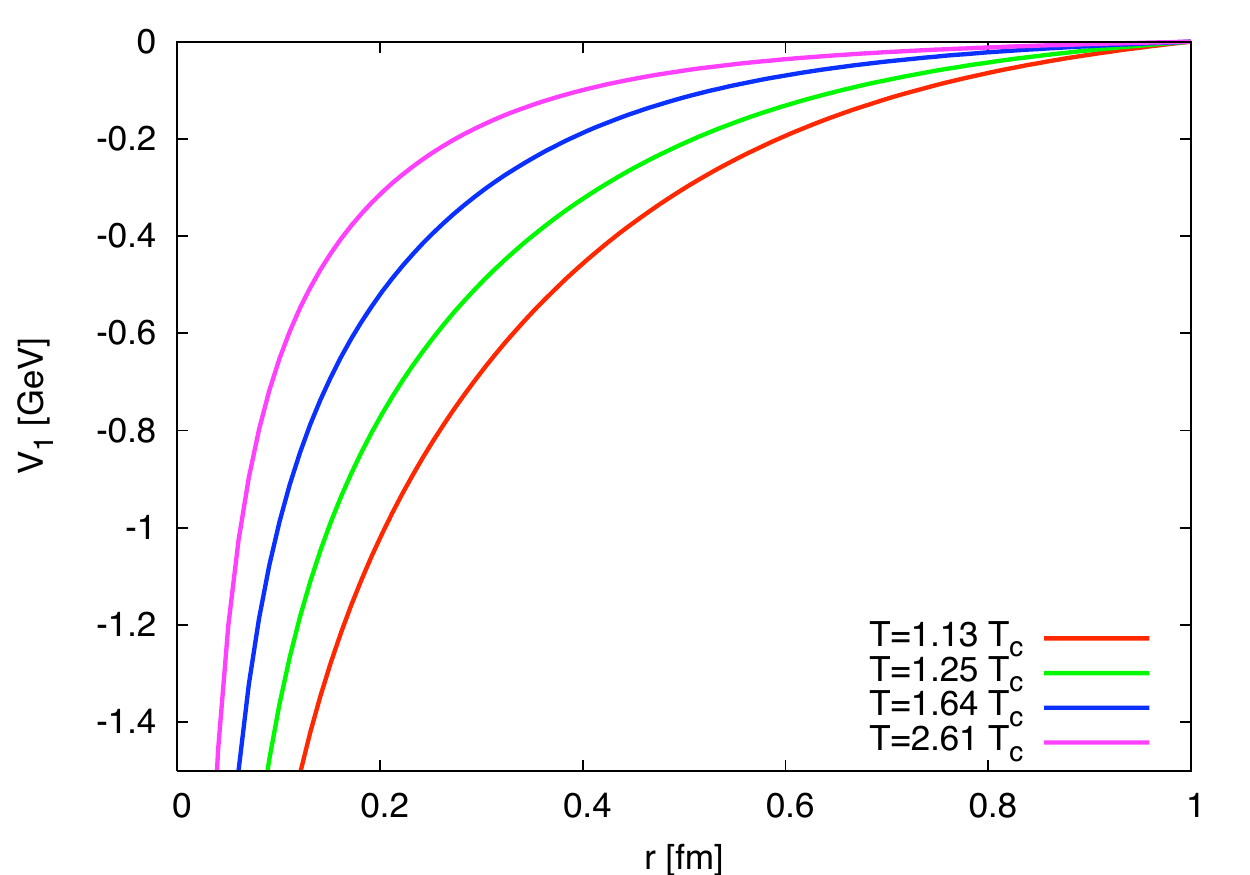}
}}
\caption{Left: Lattice results for unquenched color singlet internal energy 
compared to the parameterization in eq.~\ref{eq:param}. Right: 
Corresponding $Q\bar{Q}$ potentials from eq.~\ref{eq:pot}.}
\label{fig:potlat}
\end{figure}

Subtracting away the internal energy at $r=\infty$ is convenient when we 
use the BPL convention for the bound state projection operator $P^{'}$.
The results of the fit to the internal energy compared to the lattice data as 
well as the resulting potentials is shown in fig~\ref{fig:potlat}.

\subsection{Effective potentials from temperature-dependent Hamiltonians}

Any parametrization of the lattice results for the heavy quark potentials will be strongly 
dependent on temperature. So far, we have only considered creating effective potentials from 
temperature-independent potentials (for example, the Coulomb potential). Starting from 
temperature-dependent potentials is straightforward. However, because there seems to be 
little discussion of this in the literature, we show how our results may be generalized to the 
temperature-dependent case.

Consider the following partial differential equation,
\begin{equation}
H(\vec{r}, \beta) \rho(\vec{r}, \vec{r}^{'}, \tau) = -\frac{\partial \rho}{\partial \tau}{\rm .}
\label{eq:blochtemp}
\end{equation}
This equation is identical to the Bloch equation (eq.~\ref{eq:bloch}) but with the Hamiltonian, $H$, taken at the fixed temperature $\beta$.  Because the potential in $H$ is at a fixed temperature, the solution to \ref{eq:blochtemp} is the same as before,
\begin{equation}
\rho(\vec{r}, \vec{r}^{'}, \tau) = \left \langle \vec{r} | \exp(-\tau \hat{H}(\beta) | \vec{r}^{'} \right \rangle {\rm .}
\label{eq:denstemp}
\end{equation}
By setting $\tau=\beta$ we find that the right-hand side is exactly the density matrix.  

Therefore the prescription for a temperature-dependent Hamiltonian is simple: fix the potential at the temperature of interest for all $\tau$, then solve for the density matrix exactly as 
we have done in the previous sections ({\em i.e.} either using the Bloch equation or numerically with PIMC methods). Finally set $\tau=\beta \equiv 1/T$. This 
example shows how the analogy between the matrix elements of a density matrix, and imaginary-
time propagators, breaks down when the potential has temperature dependence.

\subsection{Results}

Using the lattice potential from the previous section the methods 
of Section \ref{veff} can be applied. Do the previous results justify using 
the Kelbg potential to approximate the effective $Q \bar Q$ potential? Well, 
since the divergent term in eq.~\ref{eq:param} is the screened Coulomb term 
at small $r$, we may work in units where $|E_1(r,T)|<\frac{1}{r}$. The 
question then is whether or not $\beta$ is sufficiently small in 
units of the Bohr energy $E_B$. The answer is yes, since in the extreme case 
where the Bohr energy is the highest and the temperature is at the lowest 
(near $T_c$), we actually have $T \sim E_B$, where our results for the 
Coulomb case clearly show the Kelbg potential working well down to 
$T=0.1E_B$. So we are confident that our methods will work well for the 
temperature range of interest.

\begin{figure}
\includegraphics[scale=.75]{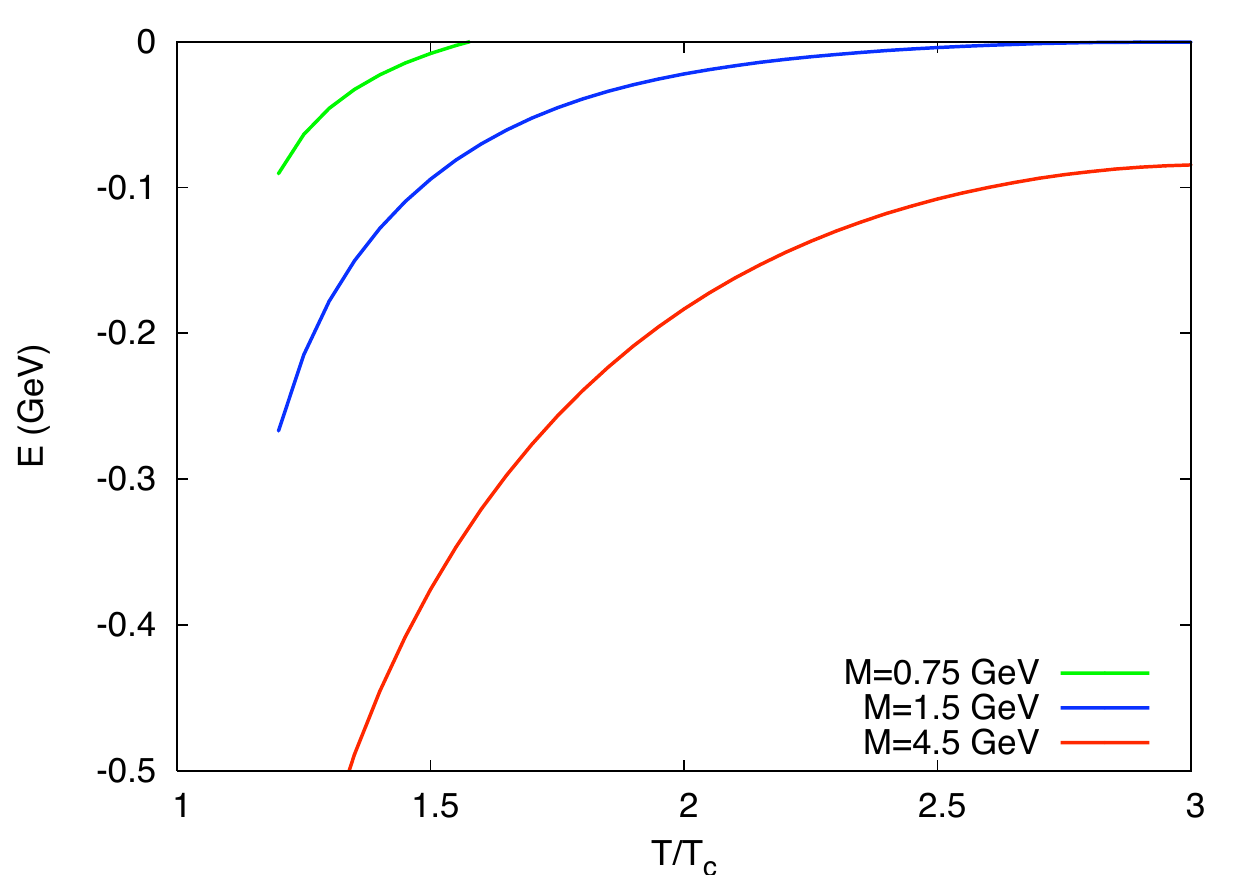}
\caption{Lowest Q\={Q} bound state energy as a function of temperature.}
\label{fig:bound}
\end{figure}

\begin{figure}
\includegraphics[scale=.75]{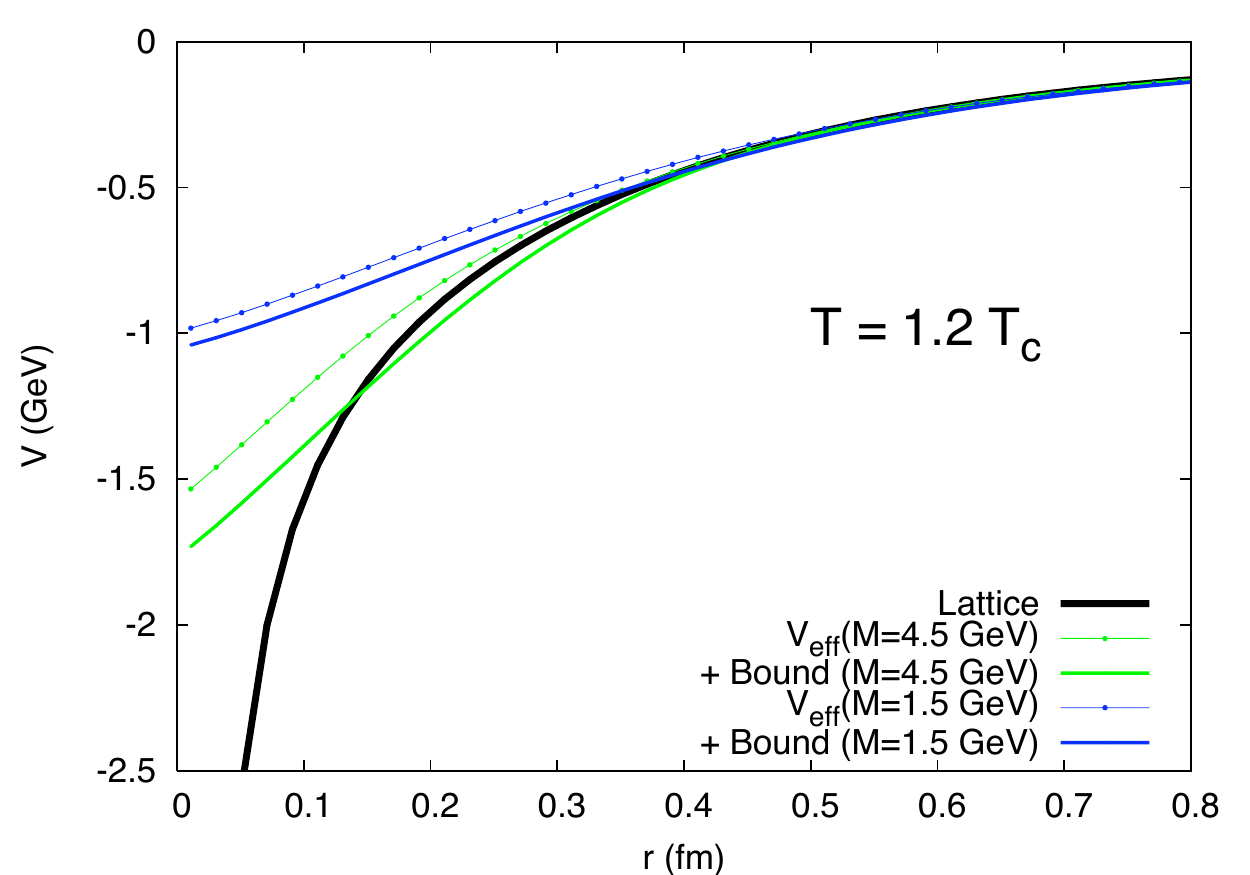}
\caption{Effective potential from the lattice data for charmonium and bottomonium at $T=1.2 T_C$.  The dotted (upper) curves show $u_1$ and the solid (lower) curves show the result after the orthogonalization to the lowest bound state.}
\label{fig:BC}
\end{figure}

In fig.~\ref{fig:BC} we show the results 
for $u_1(r)$ for charmonium and bottomonium (dotted blue and green lines 
respectively) at a temperature $T=1.2 T_c$.  As in the Coulomb case the 
effective potential takes on a finite value at $r=0$.  

It is also possible to include the effect of the lowest bound state as was 
done in the Coulomb case. The resultant energy for the lowest bound state of 
charmonium, bottomonium and also a light quark bound state having reduced 
mass $M = 0.75$ GeV is shown in fig.~\ref{fig:bound}.  Also shown in 
fig.~\ref{fig:BC} as solid lines is the effective potential when the lowest 
bound state is included using the BPL formulation.  At $T=1.2 T_c$ the 
effect of the bound state is to lower the potential by about 5\% for 
charmonium and by about 15\% for bottomonium.  The effect is even smaller at 
higher temperatures.

In order for the results given above to be useful in future simulations we now quote some useful results for Kelbg-like screened Coulomb and linear potentials.  We note that equation~\ref{eq:u1} can be integrated exactly in the case of a screened Coulomb potential, $V(r)=\frac{-\alpha e^{\mu r}}{r}$ with the result:

\begin{equation}
u_1(r,\lambda)=\frac{-\alpha}{\lambda\mu r}\sqrt{\frac{\pi}{4}}e^{-\mu r}e^{(\mu\lambda)^2/4}\times\biglb[ 1-e^{2\mu r} -2\text{erf}(\frac{\mu\lambda}{2})+\text{erf}(\frac{\mu\lambda}{2}-\frac{r}{\lambda})+e^{2\mu r}\text{erf}(\frac{\mu\lambda}{2}+\frac{r}{\lambda}) \bigrb]
\end{equation}

and for a linear potential $V(r)=br$:
\begin{equation}
u_1(r,\lambda)=\frac{b\lambda^2}{3r}\times\biglb[ 1+3\frac{r^2}{\lambda^2}-(1+\frac{r^2}{\lambda^2})e^{-r^2/\lambda^2}+\frac{\sqrt{\pi}r}{2\lambda}(3+2\frac{r^2}{\lambda^2})\biglb(1-\text{erf}(\frac{r}{\lambda})\bigrb) \bigrb]
\end{equation}

which can be used as an approximation to the {\em confinement} term in the lattice potential at small distances.  However, the inclusion of bound states which are most prominent at low temperature and higher masses must be calculated numerically from eq.~\ref{eq:bound}. 

\section{Conclusions}

We have used methods developed for treating effective potentials in electron-ion plasmas in order to generate potentials for quarkonium above deconfinement that properly take into account quantum effects and symmetry considerations.  The nature of the potential changes at small distances ($r<1\;{\rm fm}$) and reaches a finite value at zero as is also seen in the Coulomb case.  These effective potentials should be used in classical transport simulations as well as in molecular dynamic simulations of the quark-gluon plasma when $\Gamma$ is sufficiently small.


\end{document}